\hfuzz 2pt
\font\titlefont=cmbx10 scaled\magstep1
\magnification=\magstep1

\null
\vskip 1.5cm
\centerline{\titlefont ON THE WEAK-COUPLING LIMIT}
\medskip
\centerline{\titlefont AND COMPLETE POSITIVITY}
\vskip 2.5cm
\centerline{\bf F. Benatti}
\smallskip
\centerline{Dipartimento di Fisica Teorica, Universit\`a di Trieste}
\centerline{Strada Costiera 11, 34014 Trieste, Italy}
\centerline{and}
\centerline{Istituto Nazionale di Fisica Nucleare, Sezione di 
Trieste}
\vskip 1cm
\centerline{\bf R. Floreanini}
\smallskip
\centerline{Istituto Nazionale di Fisica Nucleare, Sezione di 
Trieste}
\centerline{Dipartimento di Fisica Teorica, Universit\`a di Trieste}
\centerline{Strada Costiera 11, 34014 Trieste, Italy}
\vskip 2cm
\centerline{\bf Abstract}
\smallskip
\midinsert
\narrower\narrower\noindent
We consider two non-interacting systems embedded in a heat bath.
If they remain dynamically independent,
physical inconsistencies are avoided only if the single-system 
reduced dynamics is completely positive also beyond 
the weak-coupling limit.
\endinsert
\bigskip
\vfil\eject

\noindent
{\bf 1. Introduction}
\smallskip

In a variety of different contexts, ranging from quantum optics to
the foundations of quantum mechanics, the dissipative and irreversible
time-evolutions of open quantum systems in weak interaction with 
suitable, large environments are commonly described by the so-called
quantum dynamical  semigroups~[1-3].
These consist of linear maps $\Lambda_t$, $t\geq0$, that
act on the physical states (density matrices) $\rho^S$ of the open systems,
$S$, and satisfy
an evolution equation of Kossakowski-Lindlab form~[4-10].

The maps $\Lambda_t$ are linear and completely positive, that is,
if $S$ is coupled to an arbitrary $N$-level system $S_N$, 
the maps $\Lambda_t\otimes{\rm id}_N$ preserve the positivity
of the states $\rho^{S+S_N}$ of the compound system $S+S_N$ for all $N$.

Complete positivity is an algebraic property~[10] whose
physical implications are better understood in the negative~[11]: 
if $\Lambda_t$ is
not completely positive, then for some $N$ an entangled
initial state of $S+S_N$ surely exists which develops 
negative eigenvalues under the action of
$\Lambda_t\otimes{\rm id}_N$. 
On the other hand, if the initial state of $S+S_N$ is not entangled, {\it i.e.}
$\rho^{S+S_N}=\rho^S\otimes\rho^{S_N}$, 
it never develops negative eigenvalues under the action
of $\Lambda_t\otimes{\rm id}_N$.

Physical consistency demands that the 
eigenvalues of $\Lambda_t[\rho^S]$ must be 
positive in order to be interpretable as probabilities.
It is logically necessary, but physically less compulsory, that the same 
should be true of all $\Lambda_t\otimes{\rm id}_N[\rho^{S+S_N}]$;
this latter request would only be guaranteed by
the complete positivity of $\Lambda_t$.
Yet, the system $S_N$ is totally arbitrary and unchanging, only 
statistical correlations with $S$ being allowed.
Although such an occurrence is always possible, 
it is  not always accepted~[12] as a justification~[11] of 
complete positivity as a necessary property of reduced 
dynamics.

It is sometimes argued~[12-14] that complete positivity is 
the consequence of two auxiliary 
technical simplifications that are essential in the standard~[1-3] 
derivation of quantum dynamical semigroups from the closed 
dynamics of the system $S$ plus its environment.
It is in fact assumed $a)$ that
the initial state of $S$  be
uncorrelated to that of the environment, and $b)$
that a Markov approximation is possible
on rescaled times $\tau=\lambda^2\,t$, where $\lambda<<1$ 
is the strength of the system-environment interaction~[2-9] 
(the weak-coupling limit).
However, requests $a)$ ad $b)$ 
are not always physically plausible~[12-21];
in particular, it might be necessary to
examine the subsystem dynamics on
times of the order of $\lambda^4t$, hence beyond 
the weak-coupling limit~[15, 17-20] and, in such instances, dynamics not of
completely positive type may appear.

The standard derivation of the Redfield-Bloch equations~[22], 
commonly used to describe the reduced dynamics of open two-level systems 
in chemical physics, fails to produce even positive dynamical
maps $\Lambda_t$~[4].
While the danger is fully acknowledged~[16], it is not accepted that
one should end up with a completely positive time-evolution~[4].
Rather, it is argued that only those states whose positivity is preserved 
should be physically admissible or that a slippage 
in the initial conditions is needed in order to avoid 
inconsistencies~[13-17].
However, from the previous considerations, it is clear 
that accommodating the problem of positivity does not properly 
address the issue of complete positivity which is strictly related to
quantum entanglement.

The relevance and role of complete positivity is most clearly seen
in the phenomenology of neutral K-mesons as
open quantum systems in interaction with a gravitational background.
Geometrical fluctuations at Planck's scale act as a source
of dissipation and decoherence~[23-29].
Each single K-meson is thus assumed to evolve according to a semigroup of
positivity-preserving, entropy increasing phenomenological
linear maps $\Lambda_t$.
It turns out~[25-29] that these maps must also be complete positive.
Otherwise, physical inconsistencies would plague the resulting phenomenology
of couples of K-mesons evolving in time according to the
factorized dynamical maps $\Lambda_t\otimes\Lambda_t$.
More important, these dissipative phenomenological models 
can actually be put to test in 
experiments performed at the so-called $\phi$-factories~[26,29].

In the following, we show that if the environment is such that there is
no induced interaction between two otherwise non-interacting systems 
embedded in it, then the two-system reduced dynamics is in factorized form.
Moreover this will be true at the $2$-nd and $4$-th order in the
system-environment coupling constant $\lambda$.
It then follows that if, beyond the weak-coupling limit,
the reduced dynamics is not completely positive,
then either the environment establishes a dynamical dependence
between the two subsystems or the approximations leading to
the reduced dynamics are not physically consistent.
\bigskip

\noindent
{\bf 2. Complete Positivity.}
\medskip

Let $S$ be a physical system whose time-evolution is given by a (semi)group
of linear maps $\Lambda_t$ acting on the states of $S$ represented by density
matrices $\rho^S$.
Usually, $\Lambda_t$ is positive and thus preserves the
positivity of the eigenvalues of $\rho^S$.
>From an abstract point of view~[10,11], $\Lambda_t$ is 
completely positive if and only if the map
$\Lambda_t\otimes{\rm id}_N$ acting on the states $\rho^{S+S_N}$ of
the system $S$ coupled with an arbitrary $N$-level system $S_N$, is also
positive for all possible $N$.

If $\Lambda_t$ is only positive, troubles are expected
when $S$ is coupled to a generic $N$-level system and the joint state 
$\rho^{S+S_N}$ carries correlations between $S$ and $S_N$.
In fact, if $S$ and $S_N$ are not entangled, then 
$\rho^{S+S_N}=\rho^S\otimes\rho^{S_N}$ (or a convex combinations of
factorized states), so that
$$
\Lambda_t\otimes{\rm id}_N[\rho^S\otimes\rho^{S_N}]=
\Lambda_t[\rho^S]\otimes\rho^{S_N}\ ,
\eqno(2.1)
$$
and the positivity of $\Lambda_t\otimes{\rm id}_N$ automathically follows
from the positivity of $\Lambda_t$.

However, let $S$ be a 
two-level system and $S_N=S_2=S$.
As a common state $\rho^{S+S}$ we consider
the projection 
$$
\eqalignno{
\rho_A={1\over 2}\Bigl[&\pmatrix{
1&0\cr
0&0\cr
}\otimes\pmatrix{
0&0\cr
0&1\cr
}+\pmatrix{
0&0\cr
0&1\cr
}\otimes\pmatrix{
1&0\cr
0&0\cr
}&(2.2{\rm a})\cr
-\,&\pmatrix{
0&1\cr
0&0\cr
}\otimes\pmatrix{
0&0\cr
1&0\cr
}-\pmatrix{
0&0\cr
1&0\cr
}\otimes\pmatrix{
0&1\cr
0&0\cr
}\Bigr]
&(2.2{\rm b})}
$$
onto a singlet-like state of $S+S$ with eigenvalues $0$ and $1$.
As a linear map on $S$, let us consider the transposition operation
$\displaystyle T:\,\pmatrix{a&b\cr c&d}\mapsto\pmatrix{a&c\cr b&d}$.
The map $T$ is positive, but
$$
\eqalignno{
T\otimes{\rm id}_2[\rho_A]=
{1\over 2}\Bigl[&\pmatrix{
1&0\cr
0&0\cr
}\otimes\pmatrix{
0&0\cr
0&1\cr
}+\pmatrix{
0&0\cr
0&1\cr
}\otimes\pmatrix{
1&0\cr
0&0\cr
}&(2.3{\rm a})
\cr
-&\pmatrix{
0&0\cr
1&0\cr
}\otimes\pmatrix{
0&1\cr
0&0\cr
}-\pmatrix{
0&1\cr
0&0\cr
}\otimes\pmatrix{
0&0\cr
1&0\cr
}\Bigr]&(2.3{\rm b})}
$$
has eigenvalues $\pm 1/2$.
Therefore the transposition $T$
is not completely positive, already the coupling to a $2$-level system failing
to be positive~[5].
Clearly, the origin of troubles is the term~(2.2b) which encodes 
the entanglement between the two systems
and is changed by $T\otimes{\rm id}_2$ into~(2.3b).
Instead, the term~$(2.2a)$, which represents an uncorrelated 
density matrix, is left unchanged.

Stinespring's decomposition theorem~[11] ensures that the most general 
completely positive
linear map $\Lambda$ on the states of $S$ is of the form
$$
\Lambda[\rho^S]=\sum_\ell V^\dagger_\ell\,\rho^S\,V_\ell\ ,
\eqno(2.4)
$$
where $V_\ell$ are suitable bounded operators.
Evidently, $\Lambda\otimes{\rm id}_N[\rho^{S+S_N}]$ is positive 
for any positive $\rho^{S+S_N}$.
The quantum mechanical time-evolution
$$
\Lambda_t[\rho^S]={\rm e}^{-i\,t\,H}\,\rho^S\,{\rm e}^{i\,t\,H^\dagger}
\eqno(2.5)
$$
is of the form~(2.4) and maps pure states into pure states thus preserving
coherence.
However, if one considers open quantum systems, coherence is usually lost.
Hence, $\Lambda_t$ cannot be of the form~(2.5); whether it is of the 
form~(2.4) must be decided on physical grounds.
\medskip

\noindent
{\bf Remark 2.1}\quad
In the approach to $K$-mesons as open quantum systems~[23-26], the transpositio
n
map $T$ is replaced by a phenomenological dynamical map $\Lambda_t$ and,
in~[24,26], $T\otimes{\rm id}_2$ by $\Lambda_t\otimes\Lambda_t$.
It can be shown~[26,27,29] that $\Lambda_t$ must be completely positive.
Otherwise, physically realizable initial correlated states of two K-mesons as
in~(2.2) would develop negative eigenvalues.
\bigskip

\noindent
{\bf 3. Reduced Dynamics}
\medskip

The quantum open system of interest $S$ is assumed to be weakly interacting
with a large (infinite) environment $R$, the dynamics of $S+R$ being
governed by the Hamiltonian 
$$
H=H_S+H_R+\lambda\, H_{SR}\ ,
\eqno(3.1)
$$
where $H_S$ and $H_R$ are the Hamiltonians of the system $S$, respectively
environment $R$ and $H_{SR}$ is an interaction term 
with coupling strength $\lambda$.
The total system $S+R$ is closed and its states, represented by density 
matrices $\rho$, evolve reversibly according to
$$
{\partial\rho_t\over\partial\,t}=L_H[\rho_t]:=-i\,[H\,,\,\rho_t]\ .
\eqno(3.2)
$$
The environment is supposed to be in equilibrium with respect to $H_R$, thus
$\Bigl[\rho^R\,,\,H_R\Bigr]=0$. 
Let $R$ be a heat bath in equilibrium 
at temperature $\beta^{-1}$, namely we take 
$$
\rho^R={\exp{(-\beta\,H_R)}\over{\rm Tr}\exp{(-\beta\,H_R)}}\ .
\eqno(3.3)
$$
The interaction term is chosen to be of the form
$$
H_{SR}=\sum_a\,A^S_a\otimes A^R_a\ ,
\eqno(3.4) 
$$ 
where the self-adjoint operators 
$A^S_a$ and $A^R_a$ refer to the system and
environment, respectively.
It is no restrictive to assume $\varphi_R(A^R_a)=0$ for all $a$.
\medskip

\noindent
{\bf Remarks 3.1}
\item{i)}
The environment must eventually be considered infinite dimensional
in order to allow for continuous spectra of $H_R$ and 
avoid recurrences.
The states $\rho^R$ are thus not confined to density matrices, nor are the
expectation values 
$$
\varphi_R(A^R)={\rm Tr}_R(\rho^R\,A^R)
\eqno(3.5)
$$
always expressible via a trace operation.
However, we will stick to
the density matrix notation,
the genuine case of infinitely many degrees of
freedom being easily retrieved~[5-8].
\item{ii)}
The dissipative effects perturb the time-evolution of 
$S$ given by its own Hamiltonian $H_S$ and are at least of second order
in the coupling $\lambda$.
The weak-coupling limit consists in going from the fast-time variable 
$t$ to the slow-time variable $\tau=\lambda^2t$, with $\lambda\to0$.
The technical procedure is physically justified when
the ratio $\tau_R/\tau_S$ between the characteristic time $\tau_R$ of the
environment and the characteristic time of the 
dissipative effects on $S$, $\tau_S$, is small~[3].
However, there might be heat bath temperatures for which  
one has to retain higher powers in $\tau_R/\tau_S$,
being forced to go beyond the weak-coupling limit~[15]. 
\smallskip

\noindent
In the interaction representation, $\widetilde{\rho}_t:=\exp(-tL_0)[\rho_t]$
with $L_0[\rho]:=-i\,[\,H_S+H_R\,,\,\rho\,]$, the evolution equation 
reads
$$
{\partial\widetilde{\rho}_t\over\partial\,t}=\lambda\,{\rm 
e}^{-tL_0}\,L_{H_{SR}}\,
{\rm e}^{tL_0}[\widetilde{\rho}_t]\ .
\eqno(3.6)
$$

For the sake of simplicity, we assume the spectrum of $H_S$ to be discrete and 
non-degenerate with eigenvalues $\varepsilon_r$ and eigenvectors $|r\rangle$.
We enumerate the operators $|r\rangle\langle s|$ by denoting them as $V^S_j$,
so that $H_{SR}=\sum_jV^S_j\otimes V^R_j$ with
$V^R_j=\sum_a{\rm Tr}_S(V^S_jA_a^S)A_a^R$.
Further, setting $\omega_j:=\varepsilon_r-\varepsilon_s$, we get
$$
H_{SR}(t)={\rm e}^{-tL_0}[H_{SR}]=\sum_j\,{\rm e}^{-i\omega_j\,t}\,V^S_j
\otimes {\rm e}^{i\, t\, H_R}\,V^R_j\,
{\rm e}^{-i\, t\,H_R}\ .
\eqno(3.7)
$$
\noindent
The formal solution of~(3.6) is 
$\widetilde{\rho}_t=\Bigl(1+\sum_{n=1}^\infty\lambda^n\,U^{(n)}_t\Bigr)
[\widetilde{\rho}_0]$, with
$$
U^{(n)}_t=\int_0^t{\rm d}t_1\int_0^{t_1}{\rm d}t_2
\cdots\int_0^{t_{n-1}}{\rm d}t_n\,
L_{H_{SR}(t_1)}L_{H_{SR}(t_2)}\cdots\,L_{H_{SR}(t_n)}\ ,
\eqno(3.8)
$$
where $L_A[\cdot]:=-i\Bigl[A\,,\,\cdot\Bigr]$.
  
In order to extract the system
$S$ reduced dynamics, we operate on the states $\rho$ of $S+R$
with the projector
$P[\rho]={\rm Tr_R}(\rho)\otimes\rho_R$ 
which decouples the environment degrees of freedom.
Further, we take as initial state of $S+R$ the 
state $\rho_0=\widetilde{\rho}_0=\rho^S\otimes\rho^R$ with
no correlation between $S$ and $R$.
It follows that $P[\rho_0]=\rho_0$.
\medskip

\noindent
{\bf Remark 3.2}\quad
Despite the fact that they are the most used~[1-3],
the choice of $P$ made above and the assumption on the initial global state
cannot be generically upheld.
In particular, one cannot always benefit from a factorized initial state.
However, this is in many instances plausible, as
in the case of neutral $K$-mesons in a gravitational background.
Indeed, $K$-mesons produced in strong $\phi$-meson decays
are arguably not influenced by geometrical fluctuations of gravitational origin
. 
In general, one may be forced
to adopt different projectors suited to initial states
where system $S$ and environment $R$ result correlated by 
interactions prior $t=0$~[13-17, 21].
\medskip

We now elaborate more in detail on the approach of~[19].
We assume the environment to be a Bose thermal bath described 
by the equilibrium state~(3.3).
The projector $P$ involves bath expectations with respect to~(3.3), then
only even correlation functions survive.
Keeping terms up to $\lambda^4$, one eventually finds
$$
{\partial\,P[\widetilde{\rho}_t]\over\partial\,t}
=\lambda^2\,P\dot{U}^{(2)}_tP[\widetilde{\rho}_t]\,
+\, \lambda^4\Bigl[P\dot{U}^{(4)}_t\,-\,P\dot{U}^{(2)}_tPU^{(2)}_t\Bigr]
P[\widetilde{\rho}_t]\ ,\eqno(3.9)
$$
where $\dot{U}^{(n)}_t$ is the time-derivative of $U^{(n)}$.
In particular, the $2$-nd and
$4$-th order contributions, $P\dot{U}^{(2)}_t$ and 
$P\dot{U}^{(2)}_tPU^{(2)}_t$ 
read
$$\eqalignno{
&\!\!\!\!\!\!\!\!\!
P\dot{U}^{(2)}_t=\int_0^t{\rm d}t_1\, P\,L_{H_{SR}(t)}\,L_{H_{SR}(t_1)}\ ,
&(3.10{\rm a})\cr
&\!\!\!\!\!\!\!\!\!
P\,\dot{U}^{(2)}_t\,P\,U^{(2)}_t=\int_0^t{\rm d}t_1\int_0^t{\rm d}t_2
\int_0^{t_2}{\rm d}t_3\, P\,L_{H_{SR}(t)}\,L_{H_{SR}(t_1)}\,P\,
L_{H_{SR}(t_2)}\,L_{H_{SR}(t_3)}\ .
&(3.10{\rm b})}
$$

\noindent
{\bf Remark 3.3}\quad 
After standard rearrangement of the integrals in~(3.10b), the 
whole $4$-th order contribution in~(3.9) assumes
a typical cumulant expression~[19,20].
Thermal correlation functions are expected to factorize for large 
times; in such a case the cumulants vanish and allow one to
operate a Markov approximation also at $4$-th order in $\lambda$.
\medskip

Since $\rho^R$ commutes with $H_R$, setting $\rho^S_t:={\rm Tr}_R(\rho_t)$, 
the time-evolution equation obeyed by the 
open system $S$ has the form
$$
{\partial\,\rho^S_t\over\partial\,t}=L_{H_S}[\rho^S_t]\,+\,
\lambda^2\,K^{(2)}_t[\rho^S_t]
\,+\,\lambda^4\,K^{(4)}_t[\rho^S_t]\ .
\eqno(3.11)
$$

Let $\Omega^\pm_{jk}(t):=\varphi_R\Bigl(V^R_j(\pm t)V^R_k\Bigr)$ denote
the environment two-point correlation functions. 
Then, the $2$-nd order dissipative contribution in~(3.9) explicitly reads
$$
K^{(2)}_t[\rho^S]=\sum_{j,k}\,\int_0^t{\rm d}t_1\,{\rm e}^{-i\,t_1\,\omega_j}
\Biggl(\Omega^+_{kj}(t_1)\,\Bigl[V^S_j\rho^S\,,\,V^S_k\Bigr]\, +\,
\Omega^-_{jk}(t_1)
\,\Bigl[V^S_k\,,\,\rho^S\,V^S_j\Bigr]\Biggl)\ .
\eqno(3.12)
$$

From~(3.10b) it follows that, because of the chosen projector $P$,
the $4$-th order dissipative
operator $K^{(4)}_t$ involves four-point thermal correlation functions,
which in turn are linear combinations of two-point ones.
After a lengthy calculation one arrives at
$$
K^{(4)}_t[\rho^S]=\sum_{j,k,\ell,m}\,\sum_{p=1}^{10}\,\int_0^t 
\int_0^{t-t_1}\int_0^{t-t_1-t_2}{\rm d}\vec{t}\
{\rm e}^{-i\vec{\Delta}_{k\ell m}\cdot\vec{t}}
\,\Omega_{jk\ell m}^{(p)}(\vec{t}\,)\,D_{jk\ell m}^{(p)}[\rho^S]\ ,
\eqno(3.13)
$$
where $\vec{t}=(t_1,t_2,t_3)$ and
$\vec{\Delta}_{k\ell m}\cdot\vec{t}=
(\omega_k+\omega_\ell+\omega_m)t_1+(\omega_\ell+\omega_m)t_2+\omega_mt_3$.

The quantities $\Omega^{(p)}(\vec{t}\,)$ are products of two point-correlation
functions, while the operators 
$D^{(p)}_{jk\ell m}$ are essentially double commutators of observables of $S$.
For instance, in the case $p=1$, one has
$$
\Omega_{jk\ell m}^{(1)}(\vec{t}\,):=
\Omega^+_{j\ell}(t_1+t_2)\,\Omega^+_{km}(t_2+t_3)\ ,\qquad
D_{jk\ell m}^{(1)}[\rho^S]:=\Bigl[V^S_j\,,\,\Bigl[V^S_k\,,\,V^S_\ell\Bigr]
\,V^S_m\, \rho^S\Bigr]\ .
\eqno(3.14)
$$

Because of the explicit dependence of both $K^{(2)}_t$ and $K^{(4)}_t$ on time,
the right hand side of~(3.11) retains memory effects and
does not generate a semigroup. 
However, a Markov approximation can be performed based on the following
argument (see Remark 3.3).
The two-point correlation 
functions $\varphi_R(V_j^R(t)V_k^R)$ are expected to factorize for $t$
larger than the correlation-time $\tau^R$ of
the environment which is 
much shorter than the typical time for the dissipative effects being felt by
the subsystem $S$.
Since we assumed $\varphi_R(A_a^R)=0$, it follows that
$\varphi_R(V_j^R(t)V_k^R)=0$ for $t>>\tau^R$.
Therefore, for times $t>>\tau^R$, 
the time-dependent dissipative operators $K^{(2)}_t$ and
$K^{(4)}_t$ can be replaced by time-independent
dissipative operators $K^{(2)}$ and $K^{(4)}$,
by extending to infinity each time-integration in~(3.12) and~(3.13).
For a rigorous approach to this kind of 
Markov approximation the reader is referred
to~[4,6].
Explicitly,
$$\eqalignno{
&K^{(2)}[\rho^S]:=\sum_{j,k}\,\Biggl\{
\widehat{\Omega}^+_{kj}(\omega_j)\,\Bigl[V^S_k\rho^S\,,\,V^S_j\Bigr]\, +\,
\widehat{\Omega}^-_{jk}(\omega_j)\,\Bigl[V^S_k\,,\,\rho^S\,V^S_j\Bigr]\Biggl\} 
&(3.15{\rm a})\cr
&K^{(4)}[\rho^S]:=\sum_{j,k,\ell,m}\sum_{p=1}^{10}\ ,
\widehat{\Omega}^{(p)}_{jk\ell m}(\vec{\Delta}_{k\ell m}\,)
\,D_{jk\ell m}^{(p)}[\rho^S]\ ,&(3.15{\rm b})
}$$
where
$\widehat{\Omega}_{jk}^\pm(\omega):=
\int_0^{\infty}{\rm d}t\,{\rm e}^{-i\,\omega\, t}\,\Omega^\pm_{jk}(t)$ and
$$
\widehat{\Omega}^{(p)}_{jk\ell m}(\vec{\Delta}_{k\ell m}):=
\int_0^{\infty}{\rm d}t_1\int_0^{\infty}{\rm d}t_2
\int_0^{\infty}{\rm d}t_3\,
{\rm e}^{-i\vec{\Delta}_{k\ell m}\cdot\vec{t}}\,
\Omega_{jk\ell m}^{(p)}(\vec{t}\,)\ .
\eqno(3.16)
$$
\bigskip

\noindent
{\bf 4. Non-interacting Open Quantum Systems}
\medskip

Let the open system $S$ consist 
of two non-interacting systems $S_1$ and $S_2$ whose dynamics, disregarding for
the moment the presence of the environment $R$, is governed by the
Hamiltonian operators $H_{S_1}$ and $H_{S_2}$.
Again, we assume the energies $\varepsilon_{ar}$ of $H_{S_a}$, $a=1,2$, 
to be discrete and non-degenerate
and set $\omega_{aj}:=\varepsilon_{ar}-\varepsilon_{as}$.

In absence of $R$, the system $S=S_1+S_2$ would 
evolve in time according to the Hamiltonian
$H_S=H_{S_1}\otimes{\rm id}_2\,+\,{\rm id}_1\otimes H_{S_2}$.
Instead, we suppose $S_{1,2}$ to interact weakly and independently
with a thermal bath in the state~(3.4).
We chose an interaction term of the form 
$$
H_{SR}=\sum_j\Bigl((V^S_{1j}\otimes{\rm id}_2)\otimes 
V^R_{1j}\,+\,
({\rm id}_1\otimes V^S_{2j})\otimes V^R_{2j}\Bigr)\ .
\eqno(4.1)
$$
with $\varphi_R(V^R_{aj})=0$ for all $j$, $a=1,2$,
and same coupling constant $\lambda$.

The analysis of the previous section can be repeated in
this new context, the major difference being that, inserting~(4.1) in~(3.10)
extra-indices appear identifying the system $S_a$, $a=1,2$, 
the various operators refer to.
As a consequence, the dissipative operator in~$(3.12)$ becomes
$$
K^{(2)}_t[\rho^S]=\sum_{a,b=1}^2\,\sum_{j,k}\,\int_0^t{\rm d}t_1\,
{\rm e}^{-i\omega_{aj}\,t_1}\,\Biggl\{
\Omega^+_{bk;aj}(t_1)\Bigl[V^S_{aj}\rho^S\,,\,V^S_{bk}\Bigr]\,
+\,
\Omega^-_{aj;bk}(t_1)\Bigl[V^S_{bk}\,,\,\rho^S\,V^S_{aj}
\Bigr]\Biggl\}
\eqno(4.2)
$$
where $\Omega^\pm_{aj;bk}(t)=\varphi_R\Bigl(V^R_{aj}(\pm t)
\,V^R_{bk}\Bigr)$ and, for sake of simplicity, $V^S_{aj}$ denotes either
$V^S_{1j}\otimes{\rm id}_2$ or ${\rm id}_1\otimes V^S_{2j}$.
In turn, the $4$-th order operator  reads
$$
K^{(4)}_t=\sum_{{a,b,c,d}\atop{j,k,l,m}}\sum_{p=1}^{10}
\int_0^t\int_0^{t_1}\int_0^{t-t_1-t_2}\!\!\!\!\!\!
{\rm d}\vec{t}\,
{\rm e}^{-i\vec{\Delta}_{bk;c\ell;dm}\cdot\vec{t}}\ 
\Omega^{(p)}_{aj;bk;c\ell;dm}(\vec{t}\,)
\ D_{aj;bk;c\ell;dm}^{(p)}\ ,
\eqno(4.3)
$$
where $\vec{\Delta}_{bk;c\ell;dm}\cdot \vec{t}
=(\omega_{bk}+\omega_{c\ell}+\omega_{dm})t_1\,+\,
(\omega_{c\ell}+\omega_{dm})t_2\,+\,\omega_{dm}t_3$ and
$$\eqalignno{
&\Omega_{aj;bk;c\ell;dm}^{(1)}(\vec{t}\,):=
\Omega^+_{aj;c\ell}(t_1+t_2)\,\Omega^+_{bk;dm}(t_2+t_3)&(4.4{\rm a})\cr
&D_{aj;bk;c\ell;dm}^{(1)}[\rho^S]:=\Bigl[V^S_{aj}\,,\,
\Bigl[V^S_{bk}\,,\,V^S_{c\ell}\Bigr]\,V^S_{dm}\, \rho^S\Bigr]\ .
&(4.4{\rm b})}
$$

The two-point correlation functions $\Omega^\pm_{aj;bk}(t)$
involve either
bath operators interacting with the same system, $a=c$, or  
with different systems, $a\neq b$.
We can perform the Markov
approximation exactly as in the case of just one system $S$ so that
$$\eqalignno{
&K^{(2)}[\rho^S]:=\sum_{{a,b}\atop{j,k}}\Biggl\{
\widehat{\Omega}^+_{bk;aj}(\omega_{aj})
\,\Bigl[V^S_{aj}\rho^S\,,\,V^S_{bk}\Bigr]+
\widehat{\Omega}^-_{aj;bk}(\omega_{aj})\,\Bigl[V^S_{bk}\,,\,
\rho^S\,V^S_{aj}\Bigr]\Biggl\} 
&(4.5{\rm a})\cr
&K^{(4)}[\rho^S]:=\sum_{{a,b,c,d}\atop{j,k,\ell,m}}\sum_{p=1}^{10}\,
\widehat{\Omega}^{(p)}_{aj;bk;c\ell;dm}(\vec{\Delta}_{bk;c\ell;dm}\,)
\,D_{aj;bk;c\ell;dm}^{(p)}[\rho^S]\ ,&(4.5{\rm b})
}$$
where 
$\widehat{\Omega}_{aj;bk}^\pm(\omega):=
\int_0^{\infty}{\rm d}t\,{\rm e}^{-i\,\omega\, t}\,
\Omega^\pm_{aj;bk}(t)$ and
$$
\widehat{\Omega}^{(p)}_{aj;bk;c\ell;dm}(\vec{\Delta}_{bk;c\ell;dm}):=
\int_0^{\infty}{\rm d}t_1\int_0^{\infty}{\rm d}t_2
\int_0^{\infty}{\rm d}t_3\,
{\rm e}^{-i\vec{\Delta}_{bk;c\ell;dm}\cdot\vec{t}}\,
\Omega_{aj;bk;c\ell;dm}^{(p)}(\vec{t}\,)\ .
\eqno(4.6)
$$

Using~(4.1) and~(3.7),
the $2$-nd order dissipative operator can be written as follows:
$$\eqalignno{
K^{(2)}[\rho^S]&=K_1^{(2)}\otimes{\rm id}_2[\rho^S]\,+\,
{\rm id}_1\otimes K_2^{(2)}[\rho^S]&(4.7{\rm a})\cr
&+\,\sum_{{a\neq b}\atop{j,k}}\Biggl\{
\widehat{\Omega}^+_{bk;aj}(\omega_{aj})
\,\Bigl[V^S_{aj}\rho^S\,,\,V^S_{bk}\Bigr]+
\widehat{\Omega}^-_{aj;bk}(\omega_{aj})\,\Bigl[V^S_{bk}\,,\,
\rho^S\,V^S_{aj}\Bigr]\Biggl\}\ .&(4.7{\rm b})
}
$$ 
The terms $K_a^{(2)}$, $a=1,2$ in~(4.7a) are dissipative operators of the 
form~(3.15a), 
involving only observables referring to the system $S_a$.
If the contribution~(4.7b) were absent, the right hand side of (4.7a)
would generate a factorized time-evolution
$\rho^S\mapsto\rho_t^S=\Lambda^1_t\otimes\Lambda_t^2[\rho^S]$,
with $\Lambda^1_t$ satisfying
$$
{\partial\over\partial t}\Bigl(\Lambda^1_t\otimes{\rm id}_2[\rho^S]\Bigr)
=K^{(2)}_1\otimes{\rm id}_2\Bigl[\Lambda^1_t\otimes{\rm id}_2[\rho^S]\Bigr]
\eqno(4.8)
$$
and analogously for $\Lambda^2_t$.

Clearly, the term~(4.7b) dynamically couples the two systems 
$S_1$ and $S_2$ through their interaction with the same environment.
However, this coupling depends on the strength of the thermal correlations
between bath operators describing the interaction with different subsystems.
If there are no correlatoins between them, that is if
$$
\Omega^\pm_{aj;c\ell}(t)=\varphi_R(V^R_{aj}(\pm t))\,\varphi_R(V^R_{c\ell})\ ,
\eqno(4.9)
$$
whenever $a\neq c$ and $t>0$, then the term~(4.7b) does not contribute
since, with no
restriction, we can assume the one-point bath correlation functions to vanish.

The same result follows from~(4.9) when we include $4$-th order dissipative 
effects.
This can be seen as follows.
In the expression~(4.4b), the right hand side vanishes if $b\neq c$, since
then $V^R_{bk}$ and $V^R_{c\ell}$ belong to different subsystems and thus
commute.
With $b=c$, using~(4.9) one sees that the right hand side of~(4.4a) 
vanishes unless $a=c=b=d$.
Thus, if~(4.9) holds, only $a=b=c=d=1$ and
$a=b=c=d=2$ contribute to the sum in~(4.5b).
Then the $4$-th order dissipative operator splits
as the $2$-nd order one, 
$$
K^{(4)}[\rho^S]=K^{(4)}_1\otimes{\rm id}_2[\rho^S]\,+\,
{\rm id}_1\otimes K^{(4)}_2[\rho^S]\ ,
\eqno(4.10)
$$
where
$$
K^{(4)}_a=\sum_{j,k,\ell,m}\sum_{p=1}^{10}\,
\widehat{\Omega}^{(p)}_{aj;ak;a\ell;am}(\vec{\Delta}_{ak;a\ell;am}\,)
\,D_{aj;ak;a\ell;am}^{(p)}\ ,\qquad a=1,2\ .
\eqno(4.11)
$$
As a consequence of~(4.9),
the evolution equation~(3.11), after the Markov approximation, becomes
$$
\partial_t\rho^S_t
=\Biggl(\Bigl(L_{H_{S_1}}+K^{(2)}_1+K^{(4)}_1\Bigr)\otimes{\rm id}_2
\,+\,{\rm id}_1\otimes\Bigl(L_{H_{S_2}}+K^{(2)}_2+K^{(4)}_2\Bigr)\Biggr)
[\rho^S_t]
\eqno(4.12)
$$
and the dynamical maps generated by it
factorize into $\Lambda^1_t\otimes\Lambda_t^2$, where
$\Lambda^1_t$ and $\Lambda^2_t$ are generated as in~(4.8) with $K^{(2)}_a$,
$a=1,2$, replaced by $K^{(2)}_a+K^{(4)}_a$.
\bigskip

\noindent
{\bf 5. Conclusions}
\medskip

The request that the reduced dynamics of open quantum systems in
interaction with a reservoir be completely positive is often rejected 
as not physically necessary~[12-14,16-21];
it lacks physical appeal and involves trivial and
uncontrollable couplings of the systems with generic
$N$-level systems.

The issue of complete positivity assumes its full physical
significance when dealing with the dynamics of  correlated systems 
interacting with an environment.
All depends on whether the reduced dynamics of the  
subsystems factorizes as $\Lambda_t\otimes\Lambda_t$, 
thus indicating that they evolve independently from each
other, remaining dynamically uncorrelated.

If this is the case, then the complete positivity of the single system 
reduced dynamics $\Lambda_t$ is unescapable, otherwise the joint
reduced dynamics $\Lambda_t\otimes\Lambda_t$ 
generates unacceptable negative probabilities.
>From this point of view, the necessity of complete positivity appears to be
a dynamical aspect of quantum entanglement.

On the other hand, there are situations for which the 
factorization of the reduced dynamics is not the case due to
the  subsystem-environment interaction and to
the physics of the environment itself, namely the behaviour of its
correlation functions.
If the two-system reduced dynamics does not factorize, it is not compelling
that the single-system reduced dynamics be completely positive.

In the literature there is evidence of non-completely positive
reduced dynamics beyond the weak-coupling limit~[17-20], namely taking
into account contributions of
order $\lambda^4$ in the coupling between subsystem and environment.
We have shown that if the reduced dynamics of two non-interacting subsystems
factorizes at $2$-nd order in $\lambda$, it factorizes also at 
$4$-th order.
Therefore, absence of complete positivity in the single-system
reduced dynamics at $4$-th order would jeopardize the description
of the physical behaviour of two of these
subsystems at the same order of approximation.

Finally, we would like to stress that there are physical instances where
an experimental check of complete positivity seems achievable, as in the case
of $K$-mesons~[25-29]; there, the conditions for a factorized 
reduced dynamics are plausibly fulfilled 
because of the weakness of the effects of the gravitational background.
The reduced dynamics of $K$-mesons as open quantum systems must 
then be completely positive and experiments at $\phi$-factories can explicitly
clarify this fundamental request.

\vfill\eject

\centerline{\bf REFERENCES}
\bigskip\medskip

\item{1.} E.B. Davies, {\it Quantum Theory of Open Systems}, (Academic Press,
New York, 1976)
\smallskip
\item{2.} H. Spohn, Rev. Mod. Phys. {\bf 53} (1980) 569
\smallskip
\item{3.} R. Alicki and K. Lendi, {\it Quantum Dynamical Semigroups and 
Applications}, Lect. Notes Phys. {\bf 286}, (Springer-Verlag, Berlin, 1987)
\smallskip
\item{4.} R. Dumcke and H. Spohn, Z. Physik {\bf B34} (1979) 419
\smallskip
\item{5.} V. Gorini, A. Kossakowski and
E.C.G. Surdarshan, J. Math. Phys. {\bf 17} (1976) 821 
\smallskip
\item{6.} V. Gorini, A. Frigerio, M. Verri, A. Kossakowski and
E.C.G. Surdarshan, Rep. Math. Phys. {\bf 13} (1978) 149 
\smallskip
\item{7.} V. Gorini and A. Kossakowski, J. Math. Phys. {\bf 17} (1975) 1298
\smallskip
\item{8.} A. Frigerio and V. Gorini, J. Math. Phys. {\bf 17} (1976) 2123
\smallskip
\item{9.} G. Lindblad, Commun. Math. Phys. {\bf 48} (1976) 119
\smallskip
\item{10.} M. Takesaki, {\it Theory of Operator Algebras I} (Springer, Berlin,
1979)
\smallskip
\item{11.} K. Kraus, {\it States, Effects and Operations}, Lecture Notes in
Physics 190 (Springer, Belrin, 1983)
\smallskip
\item{12.} P. Pechukas, Phys. Rev. Lett. {\bf 73} (1994) 1060
\smallskip
\item{13.} V. Romero-Rochin and I. Oppenheim, Physica {\bf A 155} (1988) 52
\smallskip
\item{14.} V. Romero-Rochin, A. Orsky and I. Oppenheim, Physica {\bf A 156} 
(1989) 244
\smallskip
\item{15.} V. Gorini, M. Verri and A. Frigerio, Physica {\bf A 161} (1989) 357
\smallskip
\item{16.} A. Suarez, R. Silbey and I. Oppenheim, J. Chem. Phys. {\bf 97} (1992
)
5101
\smallskip
\item{17.} T.-M. Chang and J.L. Skinner, Physica {\bf A 193} (1993) 483
\smallskip
\item{18.} J. Budimir and J.L. Skinner, J. Stat. Phys. {\bf 49} (1987) 1029
\smallskip
\item{19.} B.B. Laird, J. Budimir and J.L. Skinner, J. Chem. Phys. {\bf 94}
(1991) 4391
\smallskip
\item{20.} B.B. Laird and J.L. Skinner, J. Chem. Phys. {\bf 94}
(1991) 4405
\smallskip
\item{21.} A. Royer, Phys. Rev. Lett. {\bf 77} (1996) 3272
\smallskip
\item{22.} C.P. Slichter, {\it Principles of Magnetic Resonance, with Examples
from Solid State Physics} (Springer, Berlin, New York, 1990)
\smallskip
\item{23.} J. Ellis, J.S. Hagelin, D.V. Nanopoulos and M. Srednicki,
Nucl. Phys. {\bf B241} (1984) 381; 
\item{24.} P. Huet and M.E. Peskin, Nucl. Phys. {\bf B434} (1995) 3
\smallskip
\item{25.} F. Benatti and R. Floreanini, Nucl. Phys. {\bf B488} (1997) 335
\smallskip
\item{26.} F. Benatti and R. Floreanini, Nucl. Phys. {\bf B511} (1998) 550
\smallskip
\item{27.} F. Benatti and R. Floreanini, Mod. Phys. Lett. {\bf A12} (1997) 1465
\smallskip
\item{28.} F. Benatti and R. Floreanini, Banach Centre Publ. {\bf 43} (1998) 71
\smallskip
\item{29.} F. Benatti and R. Floreanini, Phys. Lett. {\bf B 468} (1999) 287

\bye